\documentclass[12pt]{article}
\usepackage{graphicx}
\usepackage{verbatim}
\usepackage{amsmath}
\usepackage{amssymb}

\def\al{\alpha}
\def\be{\beta}
\def\ga{\gamma}

\def\eps{\varepsilon}
\def\th{\theta}
\def\la{\lambda}
\def\si{\sigma}
\def\bsi{\bar\sigma}

\def\cA{{\mathcal A}}
\def\cC{{\mathcal C}}

\def\cL{{\mathcal L}}
\def\cJ{{\mathcal J}}
\def\cM{{\mathcal M}}
\def\cN{{\mathcal N}}
\def\cO{{\mathcal O}}

\def\11{{\mathbb 1}}

\def\JJ{{\mathbb J}}

\def\re{{\rm e}}
\def\ri{{\rm i}}
\def\rd{{\rm d}}

\def\tM{\tilde{M}}
\def\JJBL{\JJ_{B - L}}
\def\cJBL{\cJ_{B - L}}

\def\rX{{\rm X}}

\def\tA{\widetilde{\cA}}

\def\vev{\langle\varphi\rangle}

\def\beq{\begin{equation}}
\def\eeq{\end{equation}}
\def\bea{\begin{eqnarray}}
\def\eea{\end{eqnarray}}
\def\nn{\nonumber}

\def\ri{\text{i}}

\def\dslash{\partial \hspace{-1.3ex}\slash}

\def\pslash{p \hspace{-1.0ex}\slash}

\begin{document}
\begin{center}
{\Large\bf $(B\!-\!L)$ Symmetry {\it vs.} Neutrino Seesaw}\\[1cm]

{\bf Adam Latosi{\'n}ski$^1$, Krzysztof A. Meissner$^{1,2}$ and Hermann Nicolai$^2$}

{\it $^1$ Faculty of Physics,
University of Warsaw\\
Ho\.za 69, Warsaw, Poland\\
$^2$ Max-Planck-Institut f\"ur Gravitationsphysik
(Albert-Einstein-Institut)\\
M\"uhlenberg 1, D-14476 Potsdam, Germany\\
}

\vspace{1cm}
\begin{minipage}[t]{12cm}
{\footnotesize We compute the effective coupling of the Majoron to $W$ bosons at $\cO(\hbar)$ by evaluating the matrix element of the $(B-L)$ current between the vacuum and
a $W^+W^-$ state. The $(B\! -\! L)$ anomaly vanishes, but the amplitude does not
vanish as a result of a UV finite and non-local contribution  which is entirely due
to the mixing between left-chiral and right-chiral neutrinos. The result shows how
{\em anomaly-like} couplings may arise in spite of the fact that the $(B-L)$ current remains exactly conserved to all orders in $\hbar$, lending additional support to
our previous proposal to identify the Majoron with the axion.}
\end{minipage}
\end{center}

\vspace{0.2cm}

\section{Introduction}
The cancellation of anomalies for the $(B - L)$ current in the  Standard Model (SM)
without right-chiral neutrinos is a remarkable and well known fact (see e.g.
\cite{Bertlmann,BL} and references therein). In
this paper we consider the inclusion of right-chiral neutrinos into the SM and
demonstrate that for non-trivial  mixing between left- and right-chiral neutrinos
the relevant triangle graphs with two external electroweak vector bosons no longer
sum up to zero, as they would if only the (vanishing) $(B-L)$ anomaly were taken
into account. This result provides additional support for our previous proposal to identify
the Majoron with the axion \cite{LMN}, and clarifies some
issues that might be raised in connection with this proposal.

Recall that for a spontaneously broken abelian global symmetry,
the total Noether current $\cJ^\mu$ takes the general form
\beq\label{totalJ}
\cJ^\mu = \JJ^\mu - F_a \partial^\mu a
\eeq
where $a(x)$ is the Goldstone field and $F_a$ the parameter characterizing the scale of
spontaneous symmetry breaking, while $\JJ^\mu$ is the partial symmetry current
without spontaneous symmetry breaking. In the absence of global anomalies the total
current (\ref{totalJ}) is {\em exactly conserved to all orders in $\hbar$}. However,
the equation $\partial_\mu \cJ^\mu = 0$ says nothing about  {\em how}  the two contributions
on the r.h.s. of (\ref{totalJ}) conspire to produce overall current conservation as a
consequence of the  classical or quantum equations of motion. All it implies is that,
whenever $\Box a \neq 0$, there must be a corresponding contribution  to
$\partial_\mu \JJ^\mu\neq 0$ for (\ref{conservation}) to be satisfied, {\it viz.}
\beq\label{ME}
\Box a = F_a^{-1}  \rX \quad   \Leftrightarrow \quad \partial_\mu \JJ^\mu =  \rX
\eeq
Here the quantity
\beq\label{rX}
\rX = \rX_0 \,+\, \hbar \rX_1 \,+\, \hbar^2 \rX_2 \,+\,  \cdots
\eeq
encapsulates all (classical and quantum mechanical) contributions to the equations
of motion. While the tree level term $\rX_0$ is always local (and represents the violation
of partial current conservation $\partial_\mu \JJ^\mu$ with {\em explicit} symmetry
breaking), the quantum mechanical higher order corrections $\rX_n$ are in general
non-local. Our main point here is to show that there may arise {\em anomaly-like} contributions
in this expansion. By definition, these correspond to UV finite and non-local contributions
to the effective action that reduce to anomalous interactions
$\propto a\,  {\rm Tr}\, \cA^{\mu\nu} \tA_{\mu\nu}$ in the IR limit (where $\cA_{\mu\nu}$
can be any SM field strength). Such contributions
may appear at various orders in $\hbar$, and can mimick a non-vanishing anomaly
for topologically non-trivial gauge field configurations and constant values
of the Goldstone field.

Specifically we will be concerned with $(B-L)$ symmetry current
$\cJ^\mu = \cJBL^\mu$ and the vertex describing the coupling of the Majoron
(axion) to two external $W$ bosons at order $\hbar$. This is the simplest example
for which one can establish the existence of anomaly-like terms
in the expansion (\ref{rX}); these are entirely due to the mixing between left-chiral
and right-chiral neutrinos. The computation thus complements
our previous work \cite{LMN} where we calculated various higher loop diagrams
contributing to $\rX$ using the Yukawa interaction rather than the matrix element of
the $(B-L)$ current. In particular we derived the anomaly-like coupling of the
Majoron/axion to gluons at order $\hbar^3$. As we argued there, this
anomaly-like coupling suffices to solve the strong CP problem and therefore
removes the need for unobservable ultra-heavy new
scales, as would be required for a conventional implementation of the
Peccei-Quinn  mechanism in the SM.

Our calculation furthermore establishes the equivalence of the two field bases or
`pictures' in which the calculation of the correction terms $\rX_n$ can be
performed, and which are here related by the field redefinition (\ref{phidef}).
In one of these `pictures', the interaction occurs via a Yukawa vertex (cf. (\ref{L})
below), while in the other (cf. (\ref{Goldstone}) below) the interaction is represented
by a derivative coupling  of the Goldstone field to the $(B-L)$ current.
The first `picture' was extensively used in \cite{LMN}, whereas the calculation
based on the second `picture' adopted here closely resembles the usual anomaly
computation. With both pictures, we obtain a finite
deviation from the vanishing result expected on the basis of the vanishing $(B-L)$ anomaly. Independently of their possible relevance to axion physics the present results are thus
also of interest for the explicit determination of effective Majoron couplings which
have not been calculated in such detail in the literature.

The present work is part of a wider program in the context of the so-called Conformal
Standard Model (CSM) \cite{MN} which seeks to solve the hierarchy problem via conformal
symmetry, rather than low energy supersymmetry or large extra dimensions, by
exploiting the remarkable fact that, with the exception of the explicit mass term
in the Higgs potential, the SM is classically conformally invariant (see also
\cite{Bardeen,Hempfling,Canadians,Shap,Shap1,Foot,Ito,Lindner,Pilaftsis,Dias}
for related proposals exploiting conformality or partial conformality of the SM).
In such a framework, no intermediate scales of any kind are allowed to occur between the electroweak scale and the Planck scale. In addition smallness of couplings must be
explained via loop corrections rather than by fine-tuning explicit couplings  by hand to
very large or very small values.

\section{Lepton and baryon number  symmetry}

First we briefly recall some facts about the global symmetries of the CSM,
namely lepton and baryon number symmetry, see \cite{LMN}.
The CSM enlarges the usual SM by right-chiral neutrinos
and one extra electroweak singlet complex scalar field $\phi$. By definition, it is
classically conformally invariant because no dimensionful (mass)  parameters are
admitted in the classical Lagrangian \cite{MN,LMN}. The terms most relevant to our
discussion concern the Yukawa sector whose contribution to the CSM Lagrangian reads
\bea\label{L}
-\cL_{\rm Y}\!\!\!&=&\!\!\!
\overline{L}^i\Phi Y_{ij}^E E^j  +
\overline{Q}^i\Phi Y_{ij}^D D^j+\overline{Q}^i\eps\Phi^* Y_{ij}^U U^j\nn\\
&& + \, \overline{L}^i\eps\Phi^\ast Y_{ij}^\nu N^j+\frac12 \phi N^{i T}
\cC Y^M_{ij} N^j+{\rm h.c.}
\eea
Here the bi-spinors $Q^i$ and $L^i$ are the left-chiral quark and lepton doublets,
\beq\label{Qi}
Q^i \equiv \left(\begin{array}{c} u^i_L\\[1mm]
                                                        d^i_L \end{array}\right)\;,\quad
 L^i \equiv \left(\begin{array}{c} \nu^i_L\\[1mm]
                                                        e^i_L \end{array}\right)
\eeq
while $U^i$ and $D^i$ are the right-chiral up- and down-like quarks, $E^i$ are
the right-chiral electron-like leptons, and $N^i\equiv\nu^i_R$ the right-chiral
neutrinos (we suppress all indices except the family indices $i,j=1,2,3$).
$\Phi$ is the usual Higgs doublet, and $\phi$ is the new complex
scalar field, such that in particular all fermion mass terms are
generated by spontaneous symmetry breaking via non-vanishing expectation values
for the scalar fields and the Yukawa matrices $Y^\sharp_{ij}$.
As is evident from (\ref{L}) the electroweak singlet field $\phi$  does not directly
couple to the other SM fields, but only to right-chiral neutrinos. However, couplings
to the `observable' sector of the SM will arise through left-right neutrino mixing
and  higher loop effects.

In addition to the (local) $SU(3)_c\times SU(2)_w \times U(1)_Y$ symmetries,
the CSM Lagrangian admits two {\em global} $U(1)$ symmetries, lepton number
symmetry $U(1)_L$ and baryon number symmetry $U(1)_B$. These are, respectively,
generated by the vector-like Noether currents
\bea\label{JL}
\cJ^\mu_L &:=&  \overline{L}^i\ga^\mu L^i
+\overline{E}^i\ga^\mu E^i+\overline{N}^i\ga^\mu N^i
-2 \ri \phi^\dagger\! \stackrel{\leftrightarrow}{\partial^\mu} \!\phi      \nn\\
&\equiv&  \bar{e}^i \ga^\mu e^i + \bar\nu^i \ga^\mu \nu^i
-2 \ri \phi^\dagger\! \stackrel{\leftrightarrow}{\partial^\mu} \!\phi
\;\equiv \;\JJ_L^\mu -2 \ri \phi^\dagger\! \stackrel{\leftrightarrow}{\partial^\mu} \!\phi
\eea
and
\bea\label{JB}
\cJ^\mu_B &:=&
\frac13\,\overline{Q}^i\ga^\mu Q^i
+\frac13\,\overline{U}^i\ga^\mu U^i+\frac13\,\overline{D}^i\ga^\mu D^i\nn\\
&\equiv& \frac13 \bar{u}^i\ga^\mu u^i + \frac13 \bar{d}^i\ga^\mu d^i
\eea
where by $u^i,d^i,e^i$ and $\nu^i$ we here denote the full Dirac 4-spinors.
From (\ref{JL}) it follows that the scalar $\phi$ carries two units of lepton number
charge, hence lepton charge can `leak' from the fermions into the scalar channel.

Writing
\beq\label{phi}
\phi(x) = \varphi (x) \exp \big(\ri a(x)/\sqrt{2} \mu \big)
\eeq
we see that for $\vev\neq 0$, lepton number symmetry is spontaneously broken,
and the phase $a(x)$ becomes a Goldstone boson, the `Majoron' \cite{CMR}.  Like
$\phi(x)$, the field $a(x)$ couples only to right-chiral neutrinos at tree level,
but not to any other SM fields. For spontaneously broken lepton number symmetry
the {\em total} current $\cJ_L^\mu$ remains classically conserved, {\it i.e.}
$ \partial_\mu \cJ^\mu_L =0$, but this relation is violated at the quantum level by the
anomaly. The fermionic current $\JJ_L^\mu$ is not even conserved at the classical level.
In particular, if we replace the last term in (\ref{L}) by a Majorana mass term
\beq
\cL_{\rm Majorana} = \frac12 M_{ij} N^{iT}\cC N^j + {\rm h.c.}
\eeq
lepton number is violated explicitly, and we get
\beq\label{JL1}
\partial_\mu \JJ_L^\mu = - \ri M_{ij} N^{iT} \cC \ga^5 N^j \neq 0
\eeq
This violation of current conservation is entirely analogous to the explicit mass
dependence $\propto m_e\bar{e}\ga^5 e$ of the divergence of the axial current
$\bar{e}\ga^5\ga^\mu e$ in QED  \cite{Bertlmann}. It is
also present if the Majorana mass term is generated by spontaneous
symmetry breaking when $M_{ij} \equiv \vev Y^M_{ij}$. Using (\ref{phi}) with
$\mu = \vev =-\sqrt{2} F_a \neq 0$, the full lepton number current assumes
the universally valid form (\ref{totalJ}). Therefore the conservation of
the full current generally implies a violation of conservation for the partial fermionic
current $\JJ_L^\mu$ unless the field $a(x)$ is a free field (obeying $\Box a =0$).

While $\cJ^\mu_L$ and $\cJ^\mu_B$ are anomalous separately, the full $(B-L)$ current
\beq
\cJBL^\mu :=\cJ^\mu_B - \cJ^\mu_L \,\equiv \, \JJBL^\mu
 \, + 2 \,   \ri \phi^\dagger\! \stackrel{\leftrightarrow}{\partial^\mu} \!\phi
  \,=\,  \JJBL^\mu \, - \, \frac{\vev}{\sqrt{2}} \, \partial^\mu a
\eeq	
is quantum mechanically conserved, that is,
\beq\label{conJBL}
\partial_\mu \cJBL^\mu = 0
\eeq
to all orders in $\hbar$.  Alternatively, this relation follows by variation of the Lagrangian
\beq\label{Goldstone}
\cL_{\rm Goldstone} \, = - \frac12 \partial_\mu a\partial^\mu a
      \, + \, \frac{\sqrt{2}}{\vev} \partial_\mu a \, \JJBL^\mu
\eeq
w.r.t. to the Goldstone field $a(x)$.  Neglecting terms not relevant for this discussion,
the very same Lagrangian is obtained from the CSM Lagrangian with Yukawa
interactions (\ref{L}) by performing the $U(1)_{B-L}$ redefinition
\bea\label{phidef}
\big(L^i(x),E^i(x),N^i(x)\big)\; &\to& \;
\exp\left(-\frac{\ri a(x)}{2\sqrt{2}\mu}\right)\big(L^i(x),E^i(x),N^i(x)\big)\; , \nn\\
\big(Q^i(x),U^i(x),D^i(x)\big)\; &\to& \;
\exp\left(\frac{\ri a(x)}{6\sqrt{2}\mu}\right)\big(Q^i(x),U^i(x),D^i(x)\big)
\eea
on the fermionic fields, thereby eliminating the non-derivative Yukawa coupling of $a(x)$.
Because the $(B - L)$ current is anomaly free, the redefinition (\ref{phidef})
is in fact {\em well-defined quantum mechanically}. Therefore the change of
variables (\ref{phidef}) does not affect the fermionic functional measure, ensuring
the mutual consistency of the two formulations also at the quantum level.

Because of quantum mechanical current conservation (\ref{conJBL})
we can take up the arguments of the introduction: to satisfy the equation
\beq\label{conservation}
 \frac{\vev}{\sqrt{2}} \, \Box a =  \partial_\mu \JJBL^\mu
\eeq
with $\Box a \neq 0$, there must exist a corresponding contribution to
$\partial_\mu \JJBL^\mu$, {\it viz.}
\beq\label{ME0}
\Box a = \frac{\sqrt{2}}{\vev}\,  \rX \quad   \Leftrightarrow \quad
\partial_\mu \JJBL^\mu =  \rX
\eeq
At the classical level this claim can be easily checked by making use of the
equations of motion following from the CSM Lagrangian and  by using the fermionic
equations of motion to calculate $\partial_\mu \JJBL^\mu$. To compute the higher
order corrections in (\ref{rX}) one needs to evaluate the matrix elements
\beq\label{ME1}
\big\langle \Psi \big| a \, \partial_\mu\JJBL^\mu \big| a\big\rangle_{\rm 1PI}
\eeq
where $|\Psi\rangle$ can be any (multi-particle) state involving excitations other
than $a$, and where the subscript indicates that we amputate the external legs
in the usual fashion.

\section{Matrix elements of the leptonic current}
We  now exemplify the general arguments of the foregoing section by
determining the couplings of $a(x)$ to $W$ bosons at order $\hbar$ from (\ref{ME1}).
In  \cite{LMN} this coupling was calculated directly
from the Yukawa vertex in (\ref{L}), whereas it will be derived here from the
current coupling by evaluating the matrix element
\beq\label{JBLM0}
\big\langle W^+ W^- \big| \, \partial_\mu \JJ_{B-L}^\mu \,\big| 0 \big\rangle
\eeq
which follows from (\ref{ME1}) by factoring out the matrix element involving $a(x)$.
As for the usual anomaly this calculation reduces to the evaluation of the triangle
diagrams shown in  Fig.~1. Indeed, for the quarks the calculation is just
the standard one giving the quark contribution to the baryon number anomaly
\cite{Bertlmann}. By contrast, the leptonic contribution
is modified by the left/right neutrino mixing in such a way that the amplitude
(\ref{JBLM0}), and hence the coupling of $a(x)$ to $W$ bosons, is different from zero
for non-trivial mixing angle (whereas it would vanish without this mixing, see below).
The present calculation thus confirms our previous calculation of the axion couplings
which was based on the Yukawa Lagrangian (\ref{L}), but now in the `rotated picture'
(\ref{phidef}) where the Lagrangian assumes the form  (\ref{Goldstone}).

For simplicity, we consider only one family of leptons with right-chiral neutrinos.
Furthermore, as in our previous work, we will use
$SL(2,\mathbb{C})$ (Weyl) spinors~\footnote{Usage of this formalism is crucial
  for our calculations, whose presentation would be much more cumbersome
  in terms of  4-component spinors. For an introduction see \cite{BW}.}
to express the 4-component neutrino spinor
$\cN \equiv (\nu_L, \nu_R)\equiv (\nu_\al , \bar{N}^{\dot\al} )$ and its conjugate.
After spontaneous symmetry breaking the free part of the neutrino Lagrangian is
\bea\label{kinterms}
 \cL &=& \frac{\ri}{2}\left(\nu^{\al} \dslash_{\al\dot\be}\bar\nu^{\dot\be} +
 N^{\al} \dslash_{\al\dot\be}\bar N^{\dot\be}\right)
+ {\rm h.c.}  \nn\\
&&
- \,m\,\nu^{\al}N_\al -m\,\bar\nu_{\dot\al}\bar{N}^{\dot\al} -
\frac{M}{2} N^{\al}N_\al -\frac{M}{2}\bar{N}_{\dot\al}\bar{N}^{\dot\al}
\eea
where we have included both Dirac and Majorana mass terms, taking both
parameters real without loss of generality. As before, the fermionic lepton
number current is classically not conserved for $M\neq 0$, {\it viz.}
\beq\label{JLM}
\partial_\mu \JJ^\mu_L = \ri M \big(N^\al N_\al - \bar{N}_{\dot\al} \bar{N}^{\dot\al}\big)
\eeq

The standard procedure to deal with (\ref{kinterms}) consists in diagonalizing
the mass matrix, with rotated fields
\beq
\begin{bmatrix} \nu' \\ N' \end{bmatrix} = \begin{bmatrix} \cos \th & -\sin \th \\ \sin \th  & \cos\th \end{bmatrix} \begin{bmatrix} \nu \\ N \end{bmatrix} \label{combination}
\eeq
in terms of which the Lagrangian (\ref{kinterms}) becomes diagonal
\bea
 \cL &=& \frac{\ri}{2}\left(\nu'^{\al} \dslash_{\al\dot\be}\bar\nu'^{\dot\be} + N'^{\al} \dslash_{\al\dot\be}\bar N'^{\dot\be}\right) + {\rm c.c}    \nn\\
&&
 -\, \frac{m'}{2}\left( \nu'^{\al}\nu'_\al + \bar\nu'_{\dot\al}\bar\nu'^{\dot\al}\right) - \frac{M'}{2} \left( N'^{\al}N'_\al + \bar{N}'_{\dot\al}\bar{N}'^{\dot\al}\right)
\eea
A simple calculation gives
\beq
\tan 2\th = \frac{2m}{M} \qquad \left(0\leq \th \leq \frac{\pi}4  \;\;\;
\mbox{for} \;\; \; m\,,\,M\geq 0 \right)
\eeq
Defining the mass parameter $\tM := \sqrt{M^2 + 4m^2}$, the mass eigenvalues
are given by the seesaw formula \cite{seesaw}
\beq\label{m'}
m' = - \tM \sin^2 \th \;,\quad M' = \tM \cos^2\th
\eeq
All formulas below can then be expressed in terms of $\tM$ and the mixing angle $\th$,
and, of course,  the mass parameters of other fields. The angle $\th$ therefore interpolates
between two special limits, namely $\th =0$ when $m=0$ or $M\rightarrow\infty$ in
(\ref{kinterms}) and the right-chiral neutrino components decouple, and $\th=\pi/4$
when $m'= -M'$ [or $M=0$ in (\ref{kinterms})], and the neutrino becomes a Dirac fermion.
We note that `in real life' the value of the mixing angle $\th$ is known to be very small,
of order $10^{-6}$.

After the rotation (\ref{combination}) the propagators take the standard diagonal
form for Majorana fermions
\bea\label{prop1}
\langle \nu'_\al(x) \bar \nu'_{\dot\be}(y) \rangle &=& \ri\int\frac{d^4p}{(2\pi)^4}\frac{\pslash_{\al\dot\be} }{p^2-m'^2} \re^{-\ri p\, (x-y)} \nn\\
\langle \nu'_\al(x) \nu'_\be(y) \rangle &=&-
\ri\int\frac{d^4p}{(2\pi)^4}\frac{m' \eps_{\al\be}}{p^2-m'^2} \re^{-\ri p\, (x-y)} \\
\langle \bar\nu'_{\dot\al}(x) \bar\nu'_{\dot\be}(y) \rangle &=&-
\ri\int\frac{d^4p}{(2\pi)^4}\frac{m' \eps_{\dot\al\dot\be}}{p^2-m'^2} \re^{-\ri p\, (x-y)}\nn
\eea
with analogous expressions for the $N'$ propagators after replacing $m'\rightarrow M'$.
With the redefinitions (\ref{combination}) the SM interaction vertices now involve {\em both} neutrino components. The vertex relevant for our calculation is the one involving $W$-bosons which reads
\bea\label{LW}
\cL_{\rm int}  &=&
 -\frac{g_2}{\sqrt{2}}\,W^-_\mu \bar{e}_{L\dot\al} \bar\si^{\mu\,\dot\al \be} \big[\cos\th \,\nu'_\be +
                \sin\th \, N'_\be\big] \nn\\
    &&  -   \;   \frac{g_2}{\sqrt{2}} W^+_\mu \big[ \cos\th \,\bar{\nu}'_{\dot\al} + \sin\th \, \bar{N}'_{\dot\al}\big]
           \bsi^{\mu\,\dot\al \be} e_{L\be}
\eea
where $g_2$ is the weak coupling constant. Likewise, after the rotation (\ref{combination}), the lepton number current becomes, in terms of two-component spinors,
\bea\label{JL2}
\JJ_L^\mu
  &=& \bar{e}_{L\dot\al}\bsi^{\mu\,\dot\al\be} e_{L\be} -
  \bar{e}_{\dot\al}R\bsi^{\mu\,\dot\al \be} e_{R\be} +
\cos(2\th) \bar\nu'_{\dot\al}\bsi^{\mu\,\dot\al\be}\nu'_\be\nn
\\[2mm]
&&
- \cos(2\th) \bar{N}'_{\dot\al}\bsi^{\mu\,\dot\al \be} N'_\be \,
      + \; \sin(2\th) \big[\bar\nu_{\dot\al}'\bsi^{\mu\,\dot\al \be} N'_\be
       - \bar{N}_{\dot\al}'\bsi^{\mu\,\dot\al\be}\nu'_\be \big]
\eea

As already mentioned, we can take over the known (anomalous) result for the matrix
element $\langle W^+ W^-|\partial_\mu \JJ_B^\mu|0\rangle$, and thus need only
consider the matrix element $\langle W^+ W^-|\,\partial_\mu\JJ_L^\mu\, |0\rangle$;  at one loop this matrix element corresponds to the triangles
shown in Figure~1.\footnote{There is a similar matrix element
$\langle ZZ |\,\partial_\mu\JJ_L^\mu\, |0\rangle$  with two external $Z$-bosons, for which one of the triangles is `purely neutrino'. That calculation proceeds analogously, and with similar results, and we therefore do not discuss it here.} Due to the mixing,  there are
altogether 12 terms, which can be evaluated by standard methods (for instance,
using dimensional regularization). The final result for the amplitude is
\bea\label{cM}
-\ri\cM^{\mu\nu} (p,q) &=&
-\frac{\ri g_2^2}{16\pi^2}  \, q_\rho
\Big[ F_1\cdot \left(g^{\mu\rho}p^\nu + g^{\nu\rho}p^\mu\right) +F_2\cdot g^{\mu\nu}p^\rho \\[1mm]
&&\ \ \ \ \ \ \ \ \ \ \ \ \ \ \
 + \, F_3  \cdot \ri \eps^{\mu\nu\rho\la}p_\la  + F_4\cdot p^\mu p^\nu p^\rho/p^2 \Big] + \cO(q^2)\nn
\eea
The functions $F_i$ depend on the neutrino masses $m'$ and $M'$, as well
as on the electron mass $m_e$ and the external momentum,
\bea
F_i &=& \sin^2\th\,\cos 2\th\, K^+_i (p^2,m_e, M', M') - \sin^2 2\th\, K^+_i(p^2,m_e,m',M') \nn\\
&& -\cos^2\th\, \cos 2\th\, K^+_i(p^2,m_e,m',m')+ \sin^2\th\, K^-_i(p^2,M',m_e,m_e)\nn\\
&& + \cos^2\th\, K^-_i(p^2,m',m_e,m_e)
\eea
where $i=1,2,3,4$ and
\bea
K^{\pm}_1&=&  I_1 -I_2\pm I_3 +I_4 \nn \\
K^{\pm}_2&=&  -I_1 -I_2\mp I_3 +I_4 \nn \\
K^{\pm}_3&=&  I_1 -3I_2\pm I_3 +I_4  \nn \\
K^{\pm}_4&=&  -4I_4
\eea
The functions $I_i$ are given by  the integrals
\bea I_1(p^2,a_1,a_2,a_3) &=& \int_0^1 \rd \xi_1 \int_0^{1-\xi_1} \rd \xi_2 \log\frac{\Delta}{\mu^2} \nn \\
I_2(p^2, a_1,a_2,a_3) &=& \int_0^1 \rd \xi_1 \int_0^{1-\xi_1}\,
         \rd \xi_2\, \xi_1 \log\frac{\Delta}{\mu^2} \nn \\
I_3(p^2,a_1,a_2,a_3) &=& a_2 a_3\int_0^1 \rd \xi_1 \int_0^{1-\xi_1}  \rd \xi_2\,  \frac{1-\xi_1}{\Delta} \nn \\
I_4(p^2, a_1,a_2,a_3) &=& p^2\int_0^1 \rd \xi_1 \int_0^{1-\xi_1}  \rd \xi_2 \,
       \frac{\xi_1^2(1-\xi_1)}{\Delta} \nn
\eea
where
\beq
\Delta(p^2, a_i,\xi_i)
:= \xi_1 a_1^2 + \xi_2 a_2^2 +(1-\xi_1-\xi_2) a_3^2 - \xi_1(1-\xi_1)p^2
\eeq
and $\mu$ is a normalization parameter that drops out in the final result.
In (\ref{cM}) the coefficient function $F_3$ represents the {\em anomaly-like}
part of the amplitude, while the other coefficient functions reflect the breaking
of $SU(2)_w\times U(1)_Y$ gauge invariance.

Using symbolic algebra, all integrals can be done in closed form, but the explicit
formulae (especially for $p^2\ne 0$) are rather cumbersome, and by themselves
not very illuminating. Let us therefore concentrate on the important qualitative features.
First of all, it is easily seen that for fixed $\tilde M$ the coefficient
function $F_3$ of the anomaly-like amplitude in (\ref{cM}) varies non-trivially
with the mixing angle $\th$, and furthermore depends on the masses of the
fermions circulating in the diagram, unlike the standard
triangle anomaly \cite{Bertlmann}. Secondly, there are gauge non-invariant terms
parametrized by the functions $F_1, F_2$ and $F_4$ in (\ref{cM}),  whose presence
for generic values of $\th$ can likewise be verified numerically. Such terms are to
be expected because electroweak symmetry is broken, and the external vector
bosons are massive.

The two limiting values $\th = 0$ and $\th=\pi/4$ are special, because for them the
calculation reduces to the standard result for the anomaly of the lepton number
current in the SM. Namely, for arbitrary values of the external momentum $p^\mu$
and the mass parameters $\tM$ and $m_e$, we have
\beq\label{F124}
\lim_{\th\rightarrow\, 0\,,\frac{\pi}4} F_1 =
\lim_{\th\rightarrow 0\,,\frac{\pi}4} F_2 =
\lim_{\th\rightarrow 0\,,\frac{\pi}4} F_4 = 0
\eeq
For the anomaly-like amplitude we get
\beq\label{F3}
\lim_{\th\rightarrow 0\,,\frac{\pi}4} F_3 = \frac23
\eeq
This limit value equals the contribution from the quarks confirming the vanishing
of the amplitude (\ref{JBLM0}) for $\th=0$ and $\th = \pi/4$, in agreement with the
vanishing $(B -L)$ anomaly. The special role of these two values can be
seen as follows: for them, and only for them, the integrands of the
triangle diagrams in Fig.~1 can be re-expressed with Dirac propagators
$\propto (\gamma^\mu p_\mu + m)^{-1}$ on the internal fermion lines,
and with chiral projectors $P_L\equiv \frac12 (1 - \ga^5)$ at the $W$-vertices.
More specifically, for $\th =0$ the right-chiral component $N_\al$ decouples, and we can effectively use the massless Dirac propagator for $\nu_\al$ because of the chiral projectors $P_L$ at the vertices. For $\th=\pi/4$, on the other hand, the neutrino behaves like a massive Dirac fermion, only one chiral half of which [corresponding to the combination $(\nu'_\al + N'_\al)$] couples to the $W$ bosons in (\ref{LW}). With Dirac propagators,
it is straightforward to see that the sum of the two diagrams in Fig.~1 reduces to the difference of
two linearly divergent integrals, precisely as for the usual anomalous triangle in QED, cf. p.199 ff. in \cite{Bertlmann}.  The result is well known not to depend on the fermion masses and not to contain gauge non-invariant contributions, and is therefore the same with or without electroweak symmetry breaking, that is, proportional to
${\rm Tr}\, \cA^{\mu\nu}\tilde\cA_{\mu\nu}$ (where $\cA_{\mu\nu}$ is the
$SU(2)_w \times U(1)_Y$ Yang-Mills field strength).
In technical terms, the deviation of the result
from the customary value that we have identified here, is thus a consequence of
the fact that the neutrino propagators with $SL(2,\mathbb{C})$ spinors cannot be combined
into a Dirac propagator for a 4-spinor in the diagrams if $\th$ is different from $0$ or  $\pi/4$.

The modification of the anomaly by a finite deviation depending on $\th$  can also
be directly understood in terms of the (classical) non-conservation of the partial current
(\ref{JLM}), and using the off-diagonal neutrino propagators introduced in \cite{LMN}.
From (\ref{JLM}) we deduce
\beq\label{JBLM}
\big\langle W^+ W^- \big| \, \partial_\mu \JJ_{B-L}^\mu \,\big| 0 \big\rangle =
- \ri M \big\langle W^+ W^- \big| \big( N^\al N_\al  - \bar{N}_{\dot\al} \bar{N}^{\dot\al}\big)
| 0 \big\rangle
\eeq
Clearly, the r.h.s. vanishes if $M=0$ ($\th =\pi/4$). Less obviously, it also vanishes for
$M\rightarrow\infty$ ($\th = 0$): this is because the off-diagonal propagators converting $N$
into $\nu$ come with extra factors of $M^{-1}$  such that the matrix element of the r.h.s.
of (\ref{JBLM}) decays at least as $M^{-2}$ for large $M$. Therefore the source of the effect
is a collusion of the classical non-conservation (\ref{JLM}) and quantum mechanics:
for any state $|\Psi\rangle$ containing SM particles other than neutrinos, the matrix
elements $\langle\Psi|NN|0\rangle$ vanish at tree level, such that the non-vanishing
contributions are entirely due to loop corrections, and thus always of $\cO(\hbar)$.

\section{Conclusions}

The main result of this paper can be summarized as follows: the loop diagrams
of the fermionic $(B-L)$ current coupled to SM particles do not vanish in presence
of non-trivial mixing between left- and right-chiral neutrinos, in spite of the
vanishing $(B-L)$ anomaly. Furthermore we have shown that the phase of the
scalar field carrying lepton number charge is not a free field, and that  the amplitude obtained
for small momenta does contain anomaly-like terms. This effect depends crucially on the simultaneous presence of Dirac and Majorana mass terms, and disappears if either of
them vanishes.

\vspace{3mm}
\noindent
 {\bf Acknowledgments:} We are grateful to Wilfried Buchm\"uller and Misha Shaposhnikov
 for critical and constructive comments and discussions.

\vspace{-5mm}

\begin{figure}[ht]
\hspace{-3cm}
\begin{minipage}[b]{0.8\linewidth}
\centering
\includegraphics[scale=0.9,viewport=0 620 840 750,clip]{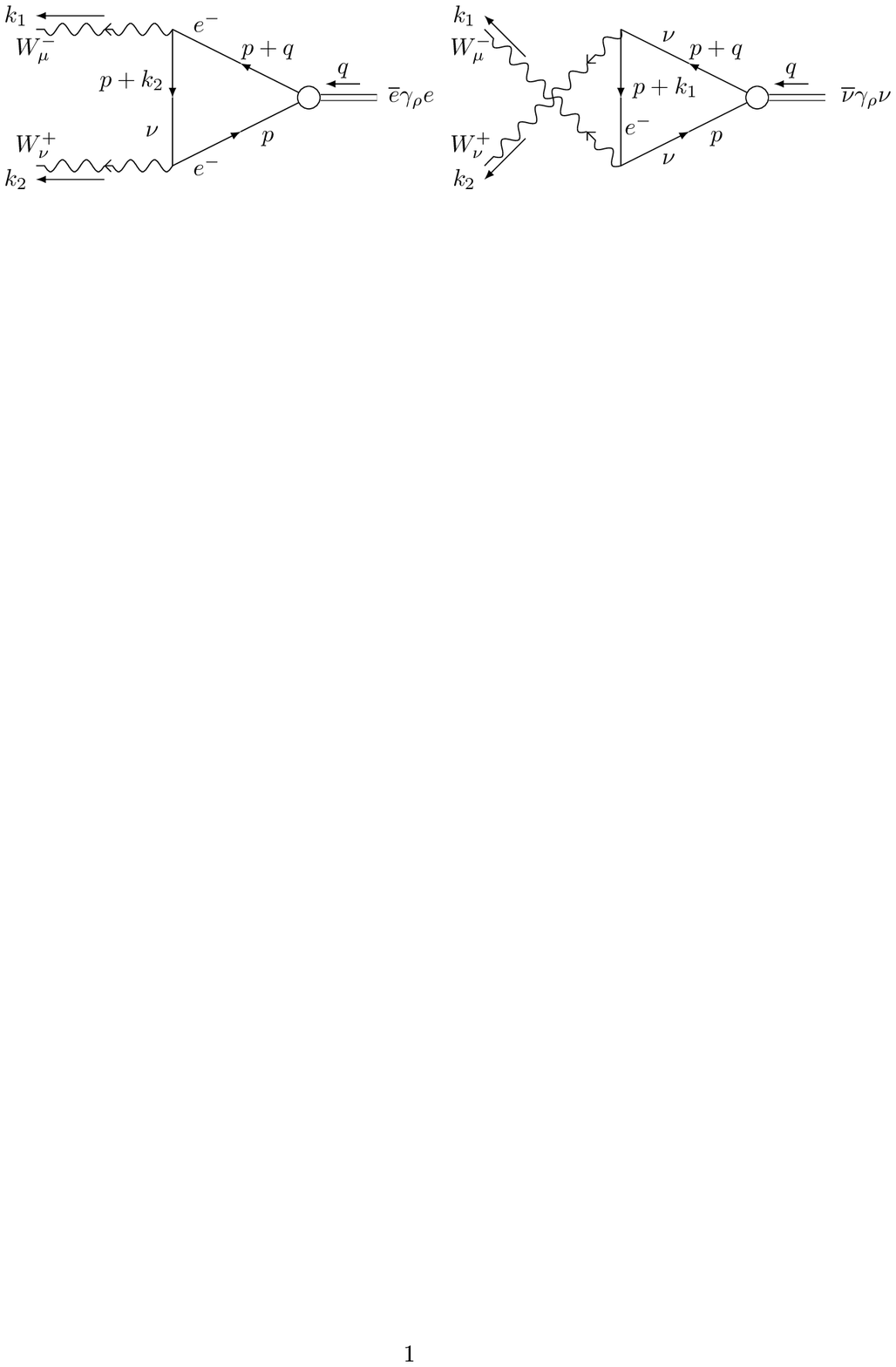}
\label{fig:figure1}
\end{minipage}
\end{figure}
\vspace{-4mm}
\begin{center}
Fig 1. Lepton number current coupling to $W$ bosons
\end{center}


\begin{thebibliography}{99}

\bibitem{Bertlmann} R.~Bertlmann, {\it Anomalies in Quantum Field Theory},
  Clarendon Press, Oxford (1996)

\bibitem{BL} W.~Buchm\"uller, R.D.~Peccei and T.~Yanagida,
     Ann. Rev. Nucl. Part. Sci. {\bf 55} (2005) 311, and references therein.

 \bibitem{LMN}
 K.A.~Meissner and H.~Nicolai, Eur.Phys. J. {\bf C 57} (2008) 493;
 A.~Latosinski, K.A.~Meissner and H.~Nicolai,
 {\tt arXiv:1203.3886[hep-th]};


\bibitem{MN} K.A.~Meissner and H.~Nicolai,
Phys. Lett. {\bf B648} (2007) 312, {\tt hep-th/0612165},

\bibitem{Bardeen} W.A.~Bardeen, {\it On Naturalness in the Standard Model},
  preprint FERMILAB-CONF-95-391-T.

\bibitem{Hempfling} R.~Hempfling,
Phys. Lett. {\bf B379} (1996) 153, hep-ph/9604278.

\bibitem{Canadians}
V. Elias, R.B. Mann, D.G.C.~McKeon and T.G.~Steele, Phys. Rev. Lett. {\bf 91} (2003) 251601;
F.A. Chishtie,V. Elias, R.B. Mann, D.G.C.~McKeon and T.G.~Steele,
Nucl. Phys. {\bf B743} (2006) 104.

\bibitem{Shap} M.~Shaposhnikov,
  arXiv:0708.3550 [hep-th] and references therein.
\bibitem{Shap1}
  M.~Shaposhnikov and F.V.~Tkachev, arXiv:0905.4857[hep-th];
  M.~Shaposhnikov and D.~Zenhausern, Phys.Lett.{\bf B671} (2009) 187,
  arXiv:0809.3406[hep-th]

\bibitem{Foot} R.~Foot, A.~Kobakhidze and R.~Volkas, Phys. Rev. {\bf D82}:035005 (2010),
 arXiv:1006.0131[hep-ph];
 R.~Foot, A.~Kobakhidze, K.L.~McDonald and R.~Volkas, Phys. Rev. {\bf D77}:035006 (2008),
arXiv:0709.2750[hep-ph]

\bibitem{Ito}S. Ito, N. Okada and Y.~Orikasa,
Phys. Lett. {\bf B676} (2009) 81, arXiv:0902.4050[hep-ph];
Phys. Rev. {\bf D80}: 115007 (2009).

\bibitem{Lindner} M.~Holthausen, M.~Lindner and M.A.~Schmidt,
    Phys. Rev. {\bf D82}:055002 (2010), arXiv:0911.0710[hep-ph];
arXiv:1112.2415[hep-ph]

\bibitem{Pilaftsis} L.~Alexander-Nunneley and A.~Pilaftsis, JHEP {\bf 1009}:021 (2010),
 arXiv:1006.5916[hep-ph].

\bibitem{Dias} A.G.~Dias and A.F.~Ferrari, Phys. Rev. {\bf D82} (2010) 085006,
{\tt arXiv:1006.5672[hep-th]}.

\bibitem{CMR} Y.~Chikashige, R.N.~Mohapatra, R.D.~Peccei,
    Phys.~Lett.~{\bf 98} (1981) 265

\bibitem{BW} J.~Bagger and J.~Wess, {\it Supersymmetry and Supergravity},
  Princeton University Press (1984).

\bibitem{seesaw}
P. Minkowski, Phys. Lett. {\bf B67} (1977) 421; T. Yanagida, Progress of Theoretical Physics 64 (1980) 1103–1105; M. Gell-Mann, P. Ramond,
and R. Slansky in Supergravity ed. by P. van
Nieuwenhuizen and D. Z. Freedman, North Holland, 1979;
S. L. Glashow, in Proc. of the 1979 Carg{\`e}se Summer Institute on Quarks and Leptons, Plenum Press 1980; R. N. Mohapatra and G. Senjanovic, Phys.Rev.Lett. {\bf 44} (1980) 912.



\end{thebibliography}
\end{document}